\documentclass[12pt]{iopart}
\usepackage{graphicx}
\usepackage{amssymb}
\begin{document}

\paper[Surface anisotropy broadening...]{Surface anisotropy broadening of the energy barrier distribution in magnetic nanoparticles}

\author{N P\'erez $^1$,  P Guardia $^1$, A G Roca $^2$, M P Morales $^2$, C J Serna $^2$, O Iglesias $^1$, F Bartolom\'e $^3$, L M Garc\'ia $^3$, X Batlle $^1$ and A Labarta $^1$}

\address{$^1$ Departament de F\'isica Fonamental and Institut de Nanoci\`encia i Nanotecnologia IN$^2$UB, Universitat de Barcelona, Mart\'i i Franqu\'es 1, 08028 Barcelona, Catalonia, Spain}

\address{$^2$ Instituto de Ciencia de Materiales de Madrid, CSIC, Sor Juana In\'es de la Cruz 3, Cantoblanco 28049, Madrid, Spain}

\address{$^3$ Instituto de Ciencia de Materiales de Arag\'on, CSIC-Universidad de Zaragoza, Dpto. F\'isica de la Materia Condensada, Pedro Cerbuna 12, 50009 Zaragoza, Spain}

\ead{nicolas@ffn.ub.es}
\date{\today}

\begin{abstract}
The effect of surface anisotropy on the distribution of energy barriers in magnetic fine particles of nanometer size is discussed within the framework of the $T\ln(t/\tau_0)$ scaling approach. The comparison between the distributions of the anisotropy energy of the particle cores, calculated by multiplying the volume distribution by the core anisotropy, and of the total anisotropy energy, deduced by deriving the master curve of the magnetic relaxation with respect to the scaling variable $T\ln(t/\tau_0)$, enables the determination of the surface anisotropy as a function of the particle size. We show that the contribution of the particle surface to the total anisotropy energy can be well described by a size--independent value of the surface energy per unit area which permits the superimposition of the distributions corresponding to the particle core and effective anisotropy energies. The method is applied to a ferrofluid composed of non-interacting Fe$_{3-x}$O$_{4}$ particles of 4.9 nm in average size and $x$ about 0.07. Even though the size distribution is quite narrow in this system, a relatively small value of the effective surface anisotropy constant $K_{s}=2.9\times 10^{-2}$ erg cm$^{-2}$ gives rise to a dramatic broadening of the total energy distribution. The reliability of the average value of the effective anisotropy constant, deduced from magnetic relaxation data, is verified by comparing it to that obtained from the analysis of the shift of the ac susceptibility peaks as a function of the frequency.
\end{abstract}

\submitto{\NT}
\maketitle

%
%
\section{\label{Introduction}Introduction}Nowadays magnetic fine particles~\cite{Batlle:2002fk} are routinely used in many technological applications such as magnetic recording~\cite{Hayashi:1996lr} and magnetic resonance imaging~\cite{Huh:2005qy, Jaeyun-Kim:2006fj}, and they are considered promising materials for various biomedical applications~\cite{Pedro-Tartaj:2003uq} and non--linear optics~\cite{ShumingNie02211997}. Consequently, the study of their magnetic properties have attracted many efforts along the last decades, in particular because considerable deviations from bulk behavior have been widely reported for particle sizes below about 100 nanometers. This is because of finite-size effects and the increasing fraction of atoms lying at the surface with lower atomic coordination than in the core as the size of the particle decreases. In fact, magnetic properties at the particle surface are governed by the breaking of the lattice symmetry associated with several chemical and physical effects leading to a site-specific surface energy, usually taken as a local uniaxial anisotropy normal to the surface. Generally, it is assumed that this local surface anisotropy averages over the whole particle surface giving rise to an effective uniaxial anisotropy acting on the net magnetization of the particle. In the nanometer range of sizes, the contribution of the surface anisotropy to the total effective anisotropy of the particle may be larger than that of the core, a fact which highly enlarges the characteristic switching time of the particle magnetization as a result of the increase in the effective energy barrier. Actually, determining the characteristic switching time by studying the non-equilibrium dynamical response of the magnetization has been one of the most used methods to get an estimation of the effective value of the particle anisotropy per unit volume, which in many cases has been found to be one or two orders of magnitude greater than that corresponding to bulk counterpart and not proportional to the particle volume~\cite{Gazeau:1998qy, gilmore:10B301, hrianca:2125, BodkerPRL1994, Respaud:1998fk}.

Many experimental results~\cite{hrianca:2125, PhysRevB.65.094409, Gazeau:1998qy, Respaud:1998fk} have been interpreted in terms of the phenomenological, ad--hoc, equation originally proposed in  Ref.~\cite{BodkerPRL1994} for the effective anisotropy per volume unit of a spherical particle of diameter $D$       
\begin{equation}
K_{eff} = K_v+\frac{6 K_s}{D},\label{1pD}
\end{equation}
where $K_v$ is the core anisotropy energy per unit volume and $K_s$ is the effective surface anisotropy per unit of surface area which, in general, is assumed to be particle--size independent. The usually adopted assumption of radial surface anisotropy is not in contradiction with equation \ref{1pD} since, in real samples, departures form ideal spherical shape and surface roughness result in an effective uniaxial contribution to $K_{s}$.

It is worth noting that for spheroidal particles, $K_{v}$ contains the contributions coming from magnetocrystalline and shape anisotropy energies. Besides $K_{eff}$ is an effective uniaxial anisotropy which represents the height of the energy barrier per unit volume blocking the swithching of the particle magnetization. $K_{v}$ and $K_{s}$ are also treated as effective  uniaxial anisotropies.

In Eq.\ \ref{1pD}, the contributions of the core and surface to the total effective anisotropy are assumed to be solely additive excluding cross--linked effects. In spite of the simplicity of this assumption and the fact that there is not theoretical justification for it, Eq.\ \ref{1pD} has been succesfully applied to show that experimental values of $K_{eff}$ determined from ac susceptibility measurements scale with $1/D$ in some fine particle systems~\cite{PhysRevB.65.094409}. Moreover, magnetic resonance experiments in maghemite nanoparticles~\cite{Gazeau:1998qy} have revealed an anisotropic contribution to the internal field associated with a positive uniaxial anisotropy originating at the particle surface which dominates over the cubic anisotropy contribution of the maghemite core and scales~\cite{Gazeau:1998qy} with $1/D$.

The ad hoc assumption of a surface anisotropy normal to the particle surface and described by a uniform surface density $K_s$ has also been applied to the study of the magnetization and switching processes of a single particle in many numerical calculations based on atomistic Monte Carlo simulations~\cite{DimitrovPhysRevB.51.11947,IglesiasPhysRevB.63.184416, KachkachiEJPB2000, kachkachi:224402}, Landau-Lifshitz-Gilbert equation~\cite{DimitrovPhysRevB.51.11947, Usatenko:2006fk} and micromagnetics models~\cite{leonov:193112}. Many of the results of these calculations reproduce most of the anomalous properties associated with surface-anisotropy effects observed in fine particle systems. In particular, a simple phenomenological model based on this assumption~\cite{leonov:193112} has been used to calculate the astroids corresponding to the phase diagrams for ellipsoidal particles which are in agreement with recent micro--SQUID experiments on isolated particles~\cite{WernsdorferPhysRevLett.78.1791}.

At the moment,  Ref.~\cite{yanes:064416} constitutes the only attempt to assess the validity of Eq.\ \ref{1pD} using an atomistic model for the surface anisotropy, namely the N\'eel model~\cite{Neel:1954fk, Gradmann:1986qy}. It has been shown that the surface energy of a particle with a cubic lattice can be effectively represented by a first order uniaxial contribution due to particle elongation, which is proportional to $K_s$ and scales with the surface; a second order contribution which is cubic in the net magnetization components, is proportional to $K_s^{2}$ and scales with the volume; and a core--surface mixing contribution which is smaller than the other two contributions. Correspondingly, the effective energy barrier of a particle could be consistent with Eq.\ \ref{1pD} only for elongated particles but not for spherical or truncated octahedral ones. However, these conclusions have been drawn in the framework of a simple atomistic model for which there is not a general justification derived from more realistic modelizations of the crystal field, spin--orbit coupling, and disorder taking place at the surface atoms.          

In this work, we show that Eq.\ \ref{1pD} may also account for the effective energy barriers of a size distribution of non-interacting spheroidal magnetic particles. We propose a method to evaluate the effective contribution of the surface and core anisotropies based on the comparison between the distributions of particle volumes, obtained from transmission electron microscopy (TEM), and energy barriers calculated from thermoremanent magnetization measurements. This method is similar to that applied in  Ref.~\cite{Silva:2007vn} to study the power law dependence of the energy barrier on the particle volume in antiferromagnetic ferrihydrite nanoparticles. We show that the effective contribution of the particle surface to the total anisotropy energy can be well described by a size--independent value of the surface anisotropy density in accordance with Eq.\ \ref{1pD}, which permits the superimposition of the two energy distributions corresponding to the particle core and total effective anisotropies. It is worth noting that relatively small values of the effective surface anisotropy density give rise to a dramatic broadening of the energy barrier distribution even for a narrow distribution of particle volumes.

%
%
\section{\label{Samples}Sample and structural characterization}Monodispersed iron oxide Fe$_{3-x}$O$_{4}$ nanoparticles were synthesized by high temperature decomposition Fe(III)-acetylacetonate, coated by oleic acid and dispersed in hexane with extra oleic acid added as stabilizer~\cite{SunS._ja026501x,RocaNanotechnology2006}. In fact, it was the sample called M5 in Ref.~\cite{Roca:2007fk}, where the full structural characterization can be found.

X-ray diffraction evidenced very good crystallinity, inverse spinel structure with lattice parameter $a=0.838(2)$ nm, similar to that of bulk magnetite, and average particle diameter $5.8\pm1.0$ nm~\cite{Roca:2007fk}.

The phase of the iron oxide particles and their stoichiometry were identified by M\"ossbauer spectroscopy~\cite{Roca:2007fk}. The spectrum recorded at 16 K was very similar to those reported for magnetite nanoparticles that have already undergone the Verwey transition~\cite{Morup:1976uq} (see Fig. 9 in Ref.~\cite{Roca:2007fk}). This spectrum was fitted to five discrete sextets following a fitting model previously proposed by other authors~\cite{Berry:1998uq,Doriguetto:2003fk} (see Table 4 in Ref.~\cite{Roca:2007fk}). Three of the five components of the spectrum, amounting 73\% of the total spectra area, showed values of the isomer shift less than 0.5 mm/s that may be attributed to Fe$^{3+}$ ions in the octahedral and tetrahedral sites of the inverse spinel structure. The other two components showed values of the isomer shift greater than 0.6 mm/s and were attributed to Fe$^{2+}$ ions lying in octahedral sites. Therefore, the Fe$^{2+}$ atomic fraction was 0.27 and the average stoichiometry of the particles was estimated to be Fe$_{2.93}$O$_{4}$. This result could be compatible with the presence of up to 21\% of maghemite phase in the form of an overoxidized shell surrounding the particle core.

\begin{figure}
\includegraphics[width=8.5cm,angle= -90]{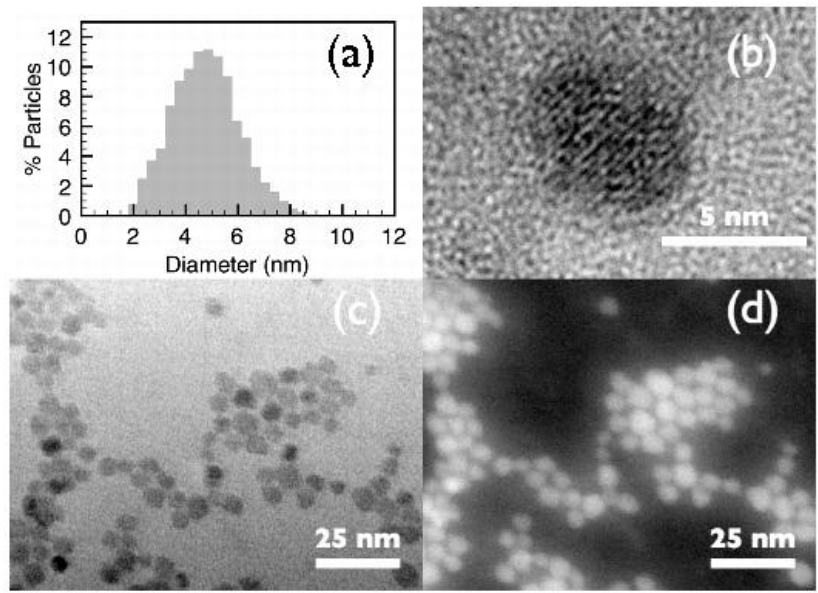}
\caption{\label{TEM}(a) Histogram of the particle diameters obtained from TEM images. (b) High resolution image of a nanoparticle showing lattice interference fringes. (c) Bright field image showing the particles. (d) Z-contrast image of the same particles as in (c).}
\end{figure}

Fig.\ \ref{TEM} shows TEM micrographs of the sample. Particles were spheroidal in shape and very uniform in size, with polydispersity lower than 20 \% of the mean size. The size distribution was determined measuring the internal diameter of about 3500 particles and the resultant hystogram was fitted to a Poisson--like distribution function, $f(D) = f_{0}D^{a}\exp(-D/b)$, with  $f_{0}=0.65$, $a=12.5$, and $b=0.36$ (in appropriate units for $D$ to be in nm). From the fitted function an average diameter $\langle D \rangle=4.9$ nm with standard deviation $\sigma=1.3$ nm was estimated.  Assuming that the particles are ellipsoidal in shape with equal minor axes, we evaluated the average value of the aspect ratio of about 100 particles to be $1.1$ with standard deviation $\sigma=0.1$. We only considered prolate shape, due to the difficulties in distinguishing prolate and oblate shapes in TEM micorgraphs.

%
%

%
%
\section{\label{Magnetic}Magnetic characterization}Magnetic measurements were carried out in a commercial Superconducting Quantum Interference Device (SQUID) magnetometer in the temperature range within 1.8 and 200 K and in magnetic fields up to 5 T. The ac susceptibility was measured by appliying an ac field of 2 Oe of amplitude and frequency within 0.1 and 1320 Hz. Magnetic relaxation at zero field and several temperatures were recorded after field cooling the sample under 50 Oe from rooom temperature down to the measuring temperature, switching off the field and then recording the magnetization decay as a function of time. All the magnetic measurements were performed using a especial container for liquid suspensions. 

Magnetization measurements revealed bulk--like saturation magnetization at 5 K, $M_{s}=78$ emu g$^{-1}$~\cite{RocaNanotechnology2006,Guardia:2007kx}.
 
\begin{figure}
\includegraphics[width=8.5cm]{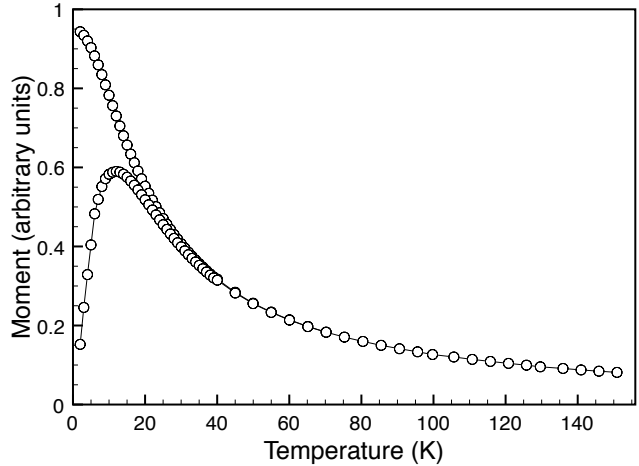}
\caption{\label{MT}Temperature dependence of the magnetization. Field cooling (upper curve) and zero field cooling (lower curve) measured under an applied magnetic field of 50 Oe. Both curves are indistinguishable for $T>T_{irr}\approx 40$ K.}
\end{figure}

The temperature dependence of the magnetization was measured increasing the temperature under an applied field of 50 Oe after zero field cooling and field cooling (ZFC--FC experiment) the sample from room temperature to 1.8 K. Fig.\ \ref{MT} shows the ZFC--FC curves of the magnetization which join togheter at $T_{irr}\approx 40$ K indicating that all the particles were superparamagnetic above this temperature. The maximum of the ZFC curve was located at a mean blocking temperature of about 15 K. The relatively small irreversibility between the ZFC and FC curves above 15 K and the abrupt increase of the FC curve below this temperature were indicative of a non-interacting or very weak interacting superparamagnetic system of nanoparticles. 

The Weiss temperature obtained extrapolating the reciprocal FC susceptibility in the superparamagnetic regime was also very small (less than 1 K) confirming a very weak strength of the interparticle interactions if they actually exist.

%
%

%
%
\section{\label{Results}Results and discussion}
%
%
\subsection{\label{Core}Core anisotropy}
M\"ossbauer spectra in Ref.~\cite{Roca:2007fk} showed that the particles undergo a Verwey transition at a certain temperature within 40 and 16 K, in accordance with previous results of other authors~\cite{Morup:1983kx} for magnetite particles of similar size. The reduced value of the Verwey temperature with respect to that of bulk magnetite may be due to finite size effects and/or the non--perfect stoichiometry of the particles. Below the temperature at which the Verwey transition takes place, magnetite is in a monoclinic configuration with uniaxial magnetic anisotropy. The effective unixial anisotropy along the [111] easy direction for stoichiometric magnetite~\cite{PhysRevB.40.9090,Borstein:2008kx} is $K_{u}\approx2.1\times10^{5}$ erg cm$^{-3}$ and almost temperature independent below 40 K. This value of the uniaxial anisotropy is also a good approximation for that of partially oxidized magnetite~\cite{PhysRevB.40.9090} with the Verwey transition occurring at a tempertaure above about 20 K, as in the case of the sample studied in this work. This is the effective bulk anisotropy that should be expected to contribute to the energy barriers blocking the particle magnetization in thermoremanent and ac magnetization experiments below about 20 K. It is worth noting that ZFC-FC curves (Fig.\ \ref{MT}) did not show any anomaly associated with the Verwey transition, likely due to smearing effects related to particle size distribution.

Taking into account that the nanoparticles are not perfect spheres we carried out a statistical evaluation of the shape anisotropy of about 100 particles from the TEM images. We assumed the particles to be ellipsoids with equal minor axes. The average value was $K_{sh}=5.3\times10^{4}$ erg cm$^{-3}$, which is one order of magnitude less than $K_{u}$. Consequently, shape anisotropy was neglected and a value $K_{v}\approx K_{u}\approx2.1\times10^{5}$ erg cm$^{-3}$ was expected.
%
%
\subsection{\label{Energy}Energy barrier distribution from magnetic relaxation}
\begin{figure}
\includegraphics[width=8.5cm, angle= -90]{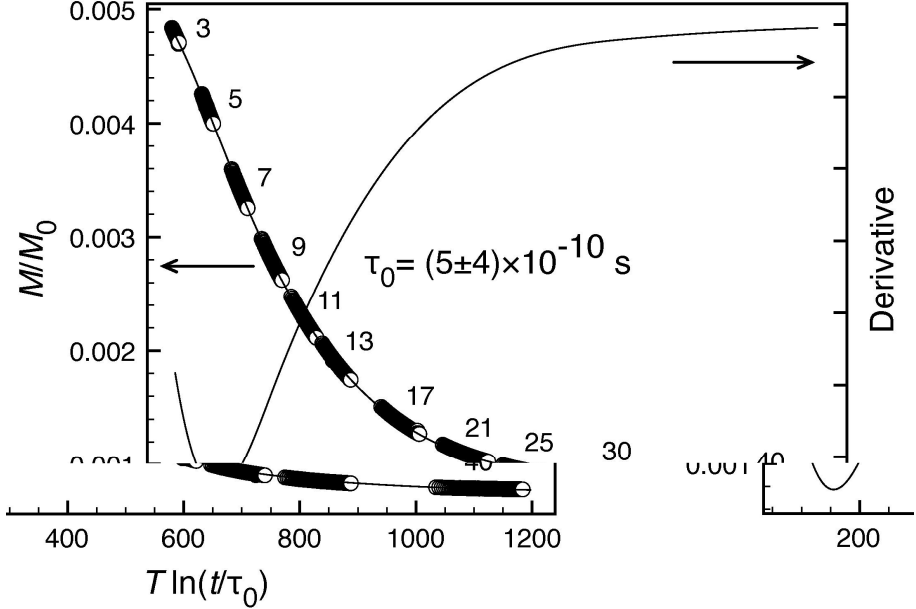}
\caption{\label{Scaling}Scaling of the relaxation curves measured at several temperatures with an attempt time of $\tau_{0}=(5\pm4)\times 10^{-10}$ s. The temperature in K corresponding to each relaxation curve is indicated beside it. Right hand side Y-axis: Derivative of the master curve with respect to the scaling variable.}
\end{figure}

In order to obtain the effective distribution of energy barriers which block the switching of the net magnetization of the particles, we measured the magnetic relaxation of the sample at constant temperature (thermoremanent magnetization) towards a demagnetized state in zero applied field after a previous cooling in the presence of 50 Oe (FC process). The obtained relaxation curves corresponding to several temperatures were plotted as a function of the scaling variable $T\ln(t/\tau_{0})$, selecting an attempt time $\tau_{0}=(5\pm4)\times 10^{-10}$ s that brought all the curves onto a single master curve~\cite{LabartaPhysRevB.48.10240}. Due to inaccuracy in the determination of the initial value of the magnetization at each temperature (the value of $M$ at $t=0$), it was also necessary to normalize the experimental data dividing them by an arbitrary reference magnetization value $M_0$ which was very close to $M_{FC}(T)$. The opposite of the derivative of the master curve with respect to $T\ln(t/\tau_{0})$ gave the distribution of energy barriers of the system~\cite{Iglesias:1996fj}. Figure~\ref{Scaling} shows the master curve for the relaxation data and its derivative as a function of the scaling variable.  When magnetic interactions among particles are very weak if present, as in the case of the sample studied in this work, the distribution of energy barriers directly represents the distribution of the effective anisotropy of the particles.
%
%

The contribution of the surface anisotropy to the effective anisotropy of the particle can be evaluated by simply comparing the  distribution of energy barriers in Fig. \ref{Scaling} with the distribution of anisotropies corresponding to the particle cores. To do that, the volume distribution was first calculated from the size distribution by transforming the particle diameter into volume assuming that the particles have spherical shape. The obtained histogram was properly renormalized by dividing the height of each bin by its width and assigning the new value of the height to the volume corresponding to the center of the bin. In the next section, we discuss a method to obtain the effective density of surface and core anisotropy energies by transforming the particle volume distribution into the total energy barrier distribution.

%
%
\subsection{\label{anisotropy}Effective surface and volume anisotropies}

According to Eq.\ \ref{1pD}, the total anisotropy energy of a single domain nanoparticle can be described in a simple model as the sum of two contributions, one proportional to its volume and another proportional to its surface area, in the form $E = K_v V+K_s S$. Assuming that the nanoparticle has spherical shape, one can rewrite the right hand of this equation in terms of the volume only, resulting in

\begin{equation}
E = K_v V+\sqrt[3]{36\pi}K_s V^{2/3}.\label{ev}
\end{equation}

In a statistical set of spherical, non-interacting nanoparticles the mean values of both sides of Eq.\ \ref{ev} have to coincide provided one uses an energy barrier distribution $f(E)$ for the left hand side and a volume distribution $g(V)$ for the right hand side. Namely
\begin{equation}
\int_{0}^{\infty}E f(E) dE =\int_{0}^{\infty}\left(K_{v}V + \sqrt[3]{36\pi}K_{s} V^{2/3} \right)  g(V) dV. \label{basica}
\end{equation}
\begin{equation}
\langle E\rangle = K_{v}\langle V\rangle + \sqrt[3]{36\pi} K_{s}\langle V^{2/3}\rangle.  \label{promedio}
\end{equation}

Calculating the derivative of Eq.\ \ref{basica} with respect to $V$, and making use of Eq. \ref{ev} to calculate $dE/dV$ and for substituting $E$ in the resulting expression, we obtain
\begin{equation}
f\left(K_v V+\sqrt[3]{36\pi}K_s V^{2/3}\right)=\left(K_{v}+2 \sqrt[3]{4\pi/(3 V)}K_{s}\right)^{-1} g(V), \label{transform2}
\end{equation}
that relates the distribution of the effective energy barriers to the particle volume distribution and enables the transformation between them. In fact, Eq. \ref{transform2} depends only on one of the anisotropy constants, provided that Eq. \ref{promedio} is used to find a relationship between $K_{s}$ and $K_{v}$. From Eq. \ref{promedio} we obtain
\begin{equation}
K_{s} = \frac{\langle E\rangle - K_{v}\langle V\rangle}{\sqrt[3]{36\pi}\langle V^{2/3}\rangle}.  \label{ks}
\end{equation}

In the previous discussion there is no need to make any assumption about the specific form of the distributions $f(E)$ and $g(V)$. In order to evaluate the averages in Eq.~\ref{ks} it is sufficient to have a conveniently large set of experimental data and carry out a numerical calculation. In our case, we fitted the experimental $f(E)$ and $g(V)$ distributions to Poisson--like distribution functions. Then, the averages in Eq.~\ref{ks} were calculated by numerical integration using the corresponding fitted functions and leading to
\begin{equation}
K_{s} = 4.9\times 10^{-2} -  8.8\times 10^{-8} K_{v} ,  \label{ksn}
\end{equation}
where $K_{s}$ and $K_{v}$ were in erg cm$^{-2}$ and  erg cm$^{-3}$, respectively.

\begin{figure}
 \includegraphics[width=8.5cm]{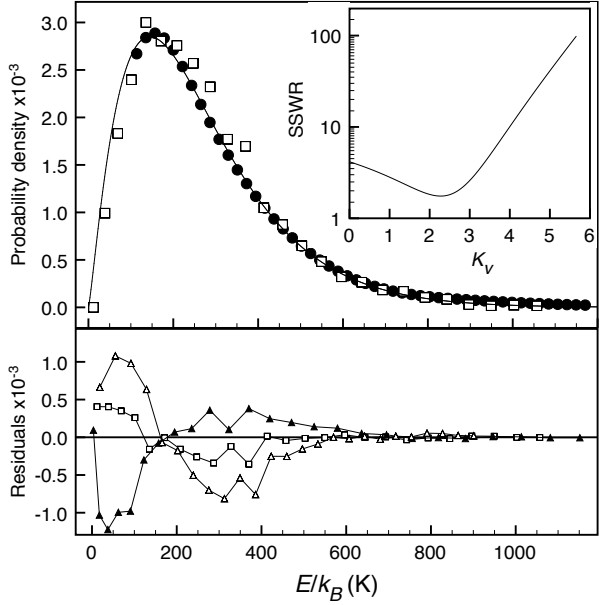}
 \caption{\label{Matching}Upper panel: Core anisotropy distribution $g(V)$ ($\Box$) transformed using Eq.\ \ref{transform2} superimposed to the energy barrier distribution $f(E)$ ($\bullet$) with $K_{v}=2.3\times 10^{5}$ erg cm$^{-3}$ and $K_{s}=2.9\times 10^{-2}$ erg cm$^{-2}$. Solid line corresponds to a Poisson--like function simultaneously fitted to both sets of experimental data. Lower panel: Residuals when fitting Eq.\ \ref{transform2} ($\Box$), imposing either $K_{s}=0$ ($\blacktriangle$) or $K_{v}=0$ ($\triangle$). Inset: Sum of squared weighted residuals of the transformed $g(V)$ from $f(E)$ using Eq.\ \ref{transform2}}
\end{figure}

A weighted least-squares fitting of the Poisson--like distribution function, corresponding to $f(E)$, to Eqs. \ref{transform2} and \ref{ksn} yielded the determination of an optimum value of $K_{v}$ that allowed the superimposition of the two distributions, $f(E)$ and $g(V)$ (see Fig.\ \ref{Matching}). Weighting of the data in the fitting procedure was done by dividing the residuals by  $f(E)$. The fitted value was $K_{v}=(2.3\pm0.7)\times 10^{5}$ erg cm$^{-3}$ from which $K_{s}=(2.9\pm0.6)\times 10^{-2}$ erg cm$^{-2}$ was estimated by using Eq.\ \ref{ksn}. The univocity of the fitting is demonstrated in the inset in Fig.\ \ref{Matching} where the sum of the squared weighted residuals shows a well-defined minimum at the fitted value of $K_{v}$. It is also worth noting that the quality of the fitting got signifincantly worse when trying to fit $f(E)$ imposing either $K_{v}=0$ or $K_{s}=0$ in equation \ref{ev} (see lower panel in Fig\ \ref{Matching} where the deviations from $f(E)$ are compared for the three cases). 
 Moreover, the fitted values $K_{s}=(25\pm2)\times10^{-2}$ erg cm$^{-2}$ and $K_{v}=(4.7\pm0.1)\times10^{5}$ erg cm$^{-3}$ obtained when imposing either $K_{v}=0$ or $K_{s}=0$ in equation \ref{ev}, respectively, were completely unphysical.
       
%
%
 
The fitted value of $K_{v}$ was in good agreement with the value expected for the core anisotropy of magnetite nanoparticles at a temperature below the Verwey transition (see Section \ref{Core}). Besides, the obtained $K_{s}$ lay within the range from 2$\times 10^{-2}$ to 6$\times 10^{-2}$ erg cm$^{-2}$ which was reported in previous experimental results for iron oxide nanoparticles~\cite{Tronc:2000lr, Gazeau:1998qy, Fiorani:1998kx, gilmore:10B301, hrianca:2125}.

In the preceeding calculations, it was implicitly assumed that $K_{s}$ is a size independent constant. This assumption seems to be supported by the good superimposition of both distributions, $f(E)$ and $g(V)$, achieved when $g(V)$ is transformed accordingly to Eq.\ \ref{transform2}. Besides, these results also confirm the applicability of Eq.\ \ref{1pD} to describe, at least in a first approximation, the effective anisotropy of the spheroidal magnetite particles studied in this work. 

\begin{figure}
\includegraphics[width=8.5cm]{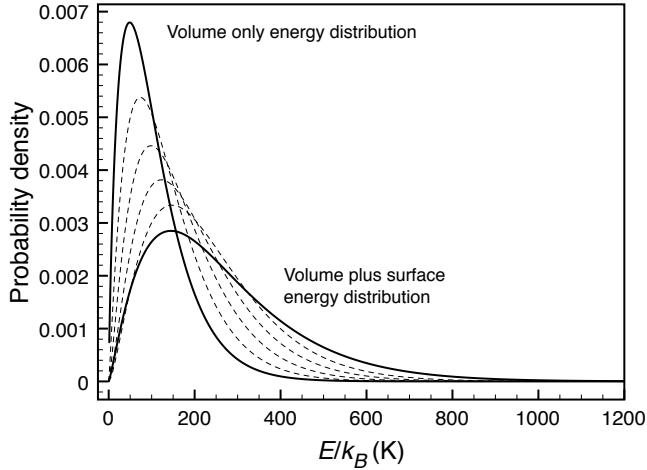}
\caption{\label{Kseffect}Poisson--like fittings to $g(V)$ and $f(E)$ (solid lines) for the magnetic nanoparticles studied in this work.  They correspond to 0\% of $K_s$ (volume only) and  100\% of $K_s$ (volume plus surface). Dashed lines correspond to the transformed $g(V)$ distribution accordingly to Eq. \ref{transform2} for 20, 40, 60, and 80 percent of $K_{s}=2.9\times 10^{-2}$ erg cm$^{-2}$ with a fixed value of $K_{v}=2.3\times10^{5}$ erg cm$^{-3}$.}
\end{figure}

Interestingly, a relatively small value of $K_s$ strongly modified $g(V)$ giving rise to a broader energy barrier distribution centered at a much higher value of the energy than that corresponding to the $g(V)$ function (core--anisotropy contribution). The strong effect of $K_s$ on $g(V)$ is emphasized in Fig.\ \ref{Kseffect} where transformed $g(V)$ distributions using Eq.\ \ref{transform2} for gradually increasing values of $K_s$ within 0 and $2.9\times 10^{-2}$ erg cm$^{-2}$ and $K_{v}=2.3\times 10^{5}$ erg cm$^{-3}$ are shown. If no surface anisotropy is present, the energy barrier distribution is identical to $g(V)$, while increasing values of $K_{s}$ gradually shift energy barriers to higher temperatures (energies) producing a significant broadening effect. 
%
%
%
   
%
%
\subsection{Effective anisotropy from ac susceptibility}The study of the blocking temperature close to the superparamagnetic regime as a function of the observational time window in ac susceptibility measurements is a conventional method to evaluate the mean value of the energy barrier which blocks switching processes of the particle magnetization. In particular, the real part of the ac susceptibility $\chi_{ac}'$ peaks at a temperture $T_{max}$ for which the particles having an energy roughly equal to the mean value of the energy barriers become blocked~\cite{DormannReview}. Therefore, at $T_{max}$, Arrhenius' law $\tau=\tau_{0}\exp[\langle E\rangle/(k_{B}T_{max})]$ is accomplished with an attempt time $\tau$ equal to the reciprocal of the frequency $f$ of the applied ac field.  Then, substituting $\tau$ by $1/f$, it can be obtained
\begin{equation}
\ln(1/f) = \ln\tau_{0} + \langle E\rangle/(k_{B}T_{max}).\label{arrhenius}
\end{equation}
Consequently, $\langle E\rangle$ can be determined by linear regression of the experimental data for $\ln(1/f)$ plotted as a function of $1/T_{max}$.

\begin{figure}
\includegraphics[width=8.5cm]{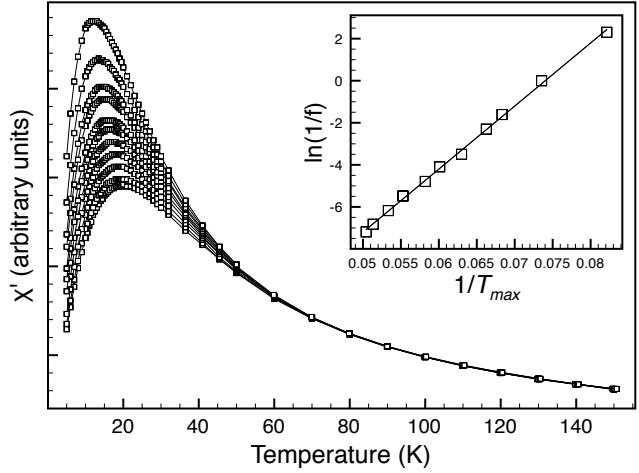}
\caption{\label{Xac}Real part of the ac susceptibility measured at frequencies (in Hz) 0.1 (upper curve), 1, 5, 10, 33, 60, 120, 240, 481, 919, and 1320 (lower curve). Inset shows $\ln(1/f)$ as a function of $1/T_{max}$. Solid line corresponds to the linear regression of the experimental points.}
\end{figure}

In Fig.\ \ref{Xac}, $\chi_{ac}'(T)$ curves for magnetite nanoparticles measured at frequencies within 0.1 and 1320 Hz are shown. The inset in Fig.\ \ref{Xac} shows $\ln(1/f)$ data as a function of $1/T_{max}$ and the corresponding linear regression, the slope of which is $\langle E\rangle/k_{B}=300$ K. This value is in good agreement with that obtained by averaging the energy barrier distribution $f(E)$ from the relaxation data which for the nanoparticles studied was $263$ K. This fact confirms the reliability of the method to obtain the energy barrier distribution associated with the effective anisotropy of the particles from magnetic relaxation analysis.

Besides, the values of the attempt time $\tau_{0}$ estimated from the ac susceptibility [$(2\pm1)\times10^{-10}$ s] and magnetic relaxation [$(5\pm4)\times10^{-10}$ s] are also in agreement taking into account the error intervals. However, it is worth noting the fact that, indeed, $\tau_{0}$ for fine particles is not a constant and shows an approximately square root dependence on the temperature~\cite{PhysRev.130.1677}. Bearing this in mind, it should be considered that the peak shifting of the real part of the ac susceptibility extended to the temperature interval from 12 to 20 K, while magnetic relaxation data scaled onto the master curve corresponded to a larger interval within 3 and 40 K. Therefore, magnetic relaxation data took into account additional switching processes at lower and higher temperatures, associated with, respectively, shorter and longer attempt times. Then, both experimental values of $\tau_{0}$ should be understood as a result of appropriate averages within the particular temperature interval where each experiment was carried out. 

%
%

%
%
\section{\label{Conclusions}Conclusions}We have proposed a method to determine the volume and surface contributions to the effective anisotropy energy in magnetic fine particles which is based on the superimposition of the distributions corresponding to volumes and energy barriers, after transformation of the volume distribution assuming a specific expression for the dependence of the energy barrier on the particle size, $E(D)$. The application of this method to a colloidal suspension of non--interacting magnetite nanoparticles of spheroidal shape has shown that the widely used expression given by Eq.\ \ref{1pD} may be a good approximation for $E(D)$ in this case.  It is stated In  Ref.~\cite{yanes:064416} that surface anisotropy in spherical particles can never produce solely an effective uniaxial anisotropy which scales with $1/D$. However, in our case, the significant uniaxial anisotropy observed may be associated with the slight deviations from the spherical shape observed in TEM micrographs and it is very plausible that the complex disorder taking place at the surface of real particles (including local modifications of the crystallographic structure, composition gradients, vacancies, dislocations and other defects) plays a key role in determining the character of their effective anisotropy since disorder may break spherical symmetry giving rise to a non--vanishing first--order uniaxial contribution to the surface anisotropy which could be dominant. Besides, magnetic frustration yielding non--collinear arrangements of the spins, modifications of the exchange interactions or the occurrence of a dead magnetic layer are phenomena having a strong potencial effect on the surface anisotropy that, in principle, cannot be excluded at the outermost shell of the particle. Moreover, the hypothetical existence of a disordered layer of finite thickness at the particles' surface would imply a reduction of the volume contributing to the core anisotropy. Thus Eq. \ref{1pD} should be rewritten in terms of the core volume and the volume of the surface shell, giving raise to a slight reduction in $K_{s}$ and an increase in $K_{v}$.

The good matching between the volume and energy barrier distributions after transformation using Eq.\ \ref{1pD} also suggests that the effective density of surface anisotropy can be considered as a size independent constant in magnetite nanoparticles, at least as a first approximation. This is contrary to the hypothesis used in  Ref.~\cite{gilmore:10B301} to obtain surface anisotropy from ac suceptibility data for magnetite nanoparticles, where the ad hoc equation $K_{s}(D)=K_s^0 \tanh (D/\lambda)$ was introduced without further justification.

It is worth noting that the actual nature of the anisotropy energy in nanoparticles may be much more complex than the Eq.\ \ref{1pD} implies, for instance, due to correlation effects between surface and core of the nanoparticles, and temperature dependence of the anisotropy constants.

However, our results suggest that Eq.\ \ref{1pD} can be used to build an effective description of the surface anisotropy contribution from the distribution of energy barriers blocking the switching of the particle magnetization, and the obtained values of the anisotropy constants should be understood as averaged over the range of temperatures in the experiments.

Finally, we would like to emphasize the strong broadening effect produced by surface anisotropy on the energy barrier distribution of magnetic fine particles even when the size distribution is quite narrow, as in the case of the sample studied in this work. This effect is an obvious consequence of the different functional dependence on the particle diameter of the energy contributions due to the core and surface anisotropy, which makes their relative importance to change dramatically as the size of the particle is reduced. This energy broadening may be relevant to give a proper interpretation of the dynamical response in systems of magnetic particles.
 
\section*{Acknowledgements}The financial support of the Spanish MEC through the projects NAN2004--08805--C04--02, NAN2004--08805--C04--01, MAT2006--03999, MAT2005--02454 and Consolider--Ingenio 2010 CSD2006--00012 is largely recognized. The Catalan DURSI (2005 SGR 00969) is also acknowledged.


\section*{References}

\begin{thebibliography}{10}

\bibitem{Batlle:2002fk}
Xavier Batlle and Am\'{i}lcar Labarta.
\newblock Finite-size effects in fine particles: magnetic and transport
  properties.
\newblock {\em J. Phys. D}, 35(6):R15--R42, 2002.

\bibitem{Hayashi:1996lr}
Takayoshi Hayashi, Shigeru Hirono, Masato Tomita, and Shigeru Umemura.
\newblock Magnetic thin films of cobalt nanocrystals encapsulated in
  graphite-like carbon.
\newblock {\em Nature}, 381(6585):772--774, 1996.

\bibitem{Huh:2005qy}
Y.~M. Huh, Y.~W. Jun, H.~T. Song, S.~Kim, J.~S. Choi, J.~H. Lee, S.~Yoon, K.~S.
  Kim, J.~S. Shin, J.~S. Suh, and J.~Cheon.
\newblock In vivo magnetic resonance detection of cancer by using
  multifunctional magnetic nanocrystals.
\newblock {\em J. Am. Chem. Soc.}, 127(35):12387--12391, 2005.

\bibitem{Jaeyun-Kim:2006fj}
Jaeyun Kim, Sungjin Park, Ji~Eun Lee, Seung~Min Jin, Jung~Hee Lee, In~Su Lee,
  Ilseung Yang, Jun-Sung Kim, Seong~Keun Kim, Myung-Haing Cho, and Taeghwan
  Hyeon.
\newblock Designed fabrication of multifunctional magnetic gold nanoshells and
  their application to magnetic resonance imaging and photothermal therapy.
\newblock {\em Angew. Chem. Int. Ed.}, 45(46):7754, 2006.

\bibitem{Pedro-Tartaj:2003uq}
Pedro Tartaj, Maria del Puerto~Morales, Sabino Veintemillas-Verdaguer, Teresita
  Gonzalez-Carreno, and Carlos~J Serna.
\newblock The preparation of magnetic nanoparticles for applications in
  biomedicine.
\newblock {\em J. Phys. D}, 36(13):R182--R197, 2003.

\bibitem{ShumingNie02211997}
Shuming Nie and Steven~R. Emory.
\newblock Probing single molecules and single nanoparticles by surface-enhanced
  raman scattering.
\newblock {\em Science}, 275(5303):1102--1106, 1997.

\bibitem{Gazeau:1998qy}
F.~Gazeau, J.~C. Bacri, F.~Gendron, R.~Perzynski, Yu.~L. Raikher, V.~I.
  Stepanov, and E.~Dubois.
\newblock Magnetic resonance of ferrite nanoparticles: evidence of surface
  effects.
\newblock {\em J. Magn. Magn. Mat.}, 186:175--187, 1998.

\bibitem{gilmore:10B301}
Keith Gilmore, Yves~U. Idzerda, Michael~T. Klem, Mark Allen, Trevor Douglas,
  and Mark Young.
\newblock Surface contribution to the anisotropy energy of spherical magnetite
  particles.
\newblock {\em J. Appl. Phys.}, 97(10):10B301, 2005.

\bibitem{hrianca:2125}
I.~Hrianca, C.~Caizer, and Z.~Schlett.
\newblock Dynamic magnetic behavior of fe$_3$o$_4$ colloidal nanoparticles.
\newblock {\em J. Appl. Phys.}, 92(4):2125--2132, 2002.

\bibitem{BodkerPRL1994}
F.~B\o{}dker, S.~M\o{}rup, and S.~Linderoth.
\newblock Surface effects in metallic iron nanoparticles.
\newblock {\em Phys. Rev. Lett.}, 72(2):282--285, Jan 1994.

\bibitem{Respaud:1998fk}
M.~Respaud, J.~M. Broto, and H.~Rakoto.
\newblock Surface effects on the magnetic properties of ultrafine cobalt
  particles.
\newblock {\em Pys. Rev. B}, 57(2):2925, 1998.

\bibitem{PhysRevB.65.094409}
F.~Luis, J.~M. Torres, L.~M. Garc\'ia, J.~Bartolom\'e, J.~Stankiewicz,
  F.~Petroff, F.~Fettar, J.-L. Maurice, and A.~Vaur\`es.
\newblock Enhancement of the magnetic anisotropy of nanometer-sized co
  clusters: Influence of the surface and of interparticle interactions.
\newblock {\em Phys. Rev. B}, 65(9):094409, 2002.

\bibitem{DimitrovPhysRevB.51.11947}
D.~A. Dimitrov and G.~M. Wysin.
\newblock Magnetic properties of spherical fcc clusters with radial surface
  anisotropy.
\newblock {\em Phys. Rev. B}, 51(17):11947--11950, May 1995.

\bibitem{IglesiasPhysRevB.63.184416}
\'Oscar Iglesias and Am\'ilcar Labarta.
\newblock Finite-size and surface effects in maghemite nanoparticles: Monte
  carlo simulations.
\newblock {\em Phys. Rev. B}, 63(18):184416, Apr 2001.

\bibitem{KachkachiEJPB2000}
H~Kachkachi, A~Ezzir, M.~Nogu{\`e}s, and E.~Tronc.
\newblock Surface effects in nanoparticles: application to maghemite
  $\gamma$-fe$_2$o$_3$.
\newblock {\em Eur. Phys. J. B}, 14:681--689, 2000.

\bibitem{kachkachi:224402}
H.~Kachkachi and E.~Bonet.
\newblock Surface-induced cubic anisotropy in nanomagnets.
\newblock {\em Phys. Rev. B}, 73(22):224402, 2006.

\bibitem{Usatenko:2006fk}
O.V. Usatenko, O.A. Chubykalo-Fesenko, and F.G. Sanchez.
\newblock Nonlinear adiabatic dynamics of small ferromagnetic particles.
\newblock {\em Int. J. Mod. Phys. B}, 20:5391--5404, 2006.

\bibitem{leonov:193112}
A.~A. Leonov, I.~E. Dragunov, and A.~N. Bogdanov.
\newblock Surface-induced anisotropy and multiple states in elongated magnetic
  nanoparticles.
\newblock {\em Appl. Phys. Lett}, 90(19):193112, 2007.

\bibitem{WernsdorferPhysRevLett.78.1791}
W.~Wernsdorfer, E.~Bonet Orozco, K.~Hasselbach, A.~Benoit, B.~Barbara,
  N.~Demoncy, A.~Loiseau, H.~Pascard, and D.~Mailly.
\newblock Experimental evidence of the n\'eel-brown model of magnetization
  reversal.
\newblock {\em Phys. Rev. Lett.}, 78(9):1791--1794, Mar 1997.

\bibitem{yanes:064416}
R.~Yanes, O.~Chubykalo-Fesenko, H.~Kachkachi, D.~A. Garanin, R.~Evans, and
  R.~W. Chantrell.
\newblock Effective anisotropies and energy barriers of magnetic nanoparticles
  with n\'eel surface anisotropy.
\newblock {\em Phys. Rev. B}, 76(6):064416, 2007.

\bibitem{Neel:1954fk}
Louis N\'eel.
\newblock Magnetic surface anisotropy and superlattice formation by
  orientation.
\newblock {\em J. Phys. Rad.}, 15:376, 1954.

\bibitem{Gradmann:1986qy}
Ulrich Gradmann.
\newblock Magnetic surface anisotropies.
\newblock {\em J. Magn. Magn. Mat.}, 54-57:733--736, 1986.

\bibitem{Silva:2007vn}
N.J.O. Silva, V.S Amaral, L.~D. Carlos, B.~Rodr{\'\i}guez-Gonz{\'a}lez, L.~M.
  Liz-Marz{\'a}n, T.~S. Berqu{\'o}, S.~K. Benarjee, V.~de~Zea~Bermudez,
  A.~Mill{\'a}n, and F.~Palacio.
\newblock Evidence of random magnetic anisotropy in ferrhydrite nanoparticles
  basen on analysis of statistical distributions.
\newblock {\em Phys. Rev. B}, 77:134426, 2008.

\bibitem{SunS._ja026501x}
S.~Sun and H.~Zeng.
\newblock Size-controlled synthesis of magnetite nanoparticles.
\newblock {\em J. Am. Chem. Soc.}, 124(28):8204--8205, 2002.

\bibitem{RocaNanotechnology2006}
A~G Roca, M~P Morales, K~O'Grady, and C~J Serna.
\newblock Structural and magnetic properties of uniform magnetite nanoparticles
  prepared by high temperature decomposition of organic precursors.
\newblock {\em Nanotechnology}, 17(11):2783--2788, 2006.

\bibitem{Roca:2007fk}
A.~G. Roca, J.~F. Marco, M.~P. Morales, and C.~J. Serna.
\newblock Effect of naure and particle size on properties of uniform magnetite
  and magnemite nanoparticles.
\newblock {\em J. Phys. Chem C}, 111:18577, 2007.

\bibitem{Morup:1976uq}
S.~M\o{}rup, h~Tops\o{}e, and J~Lipka.
\newblock Modifed theory for m{\"o}ssbauer spectra of superparamagnetic
  particles: application to fe$_{3}$o$_{4}$.
\newblock {\em J. Phys. Coll.}, (C6):287, 1976.

\bibitem{Berry:1998uq}
F.~J. Berry, S.~Skinner, and M.~F. Thomas.
\newblock {\em J. Phys. Condens. Matter}, 10:215--220, 1998.

\bibitem{Doriguetto:2003fk}
A.~C. Doriguetto, N.~G. Fernandes, E~Persiano, A. I. C. Numes~Filho, J.~M.
  Greneche, and J.~D. Fabris.
\newblock {\em Phys. Chem. Miner.}, 30:249, 2003.

\bibitem{Guardia:2007kx}
P.~Guardia, B.~Batlle-Brugal, A.~G. Roca, O.~Iglesias, M.~P. Morales, C.~J.
  Serna, A.~Labarta, and X.~Batlle.
\newblock Surfactant effects in magnetite nanoparticles of controlled size.
\newblock {\em J. Magn. Magn. Mat.}, 316:e756--e759, 2007.

\bibitem{Morup:1983kx}
S.~M\o{}rup.
\newblock {\em J. Magn. Magn. Mat.}, 39:45, 1983.

\bibitem{PhysRevB.40.9090}
Z.~Ka\ifmmode~\mbox{\c{}}\else \c{}\fi{}kol and J.~M. Honig.
\newblock Influence of deviations from ideal stoichiometry on the anisotropy
  parameters of magnetite fe$_{3(1-\delta)}$o$_4$.
\newblock {\em Phys. Rev. B}, 40(13):9090--9097, Nov 1989.

\bibitem{Borstein:2008kx}
R.A. Lefever.
\newblock Magnetic and other properties of oxides and related compounds:
  Spinels, fe oxides, and fe-me-o compounds.
\newblock In K.~H. Hellwege and A.~M. Hellwege, editors, {\em
  Landolt--B{\"o}rnstein Numerical Data and Functional Relationships in Science
  and Technology, New Series}, volume III/12b, pages 61, 62. Springer Verlag,
  1980.

\bibitem{LabartaPhysRevB.48.10240}
A.~Labarta, O.~Iglesias, Ll. Balcells, and F.~Badia.
\newblock Magnetic relaxation in small-particle systems: $\ln(t/\tau_0)$
  scaling.
\newblock {\em Phys. Rev. B}, 48(14):10240--10246, Oct 1993.

\bibitem{Iglesias:1996fj}
O.~Iglesias, F.~Badia, A.~Labarta, and Ll. Balcells.
\newblock Energy barrier distributions in magnetic systems from the
  $t\ln(t/\tau_0)$ scaling.
\newblock {\em Z. Phys. B}, 100:173--178, 1996.

\bibitem{Tronc:2000lr}
E.~Tronc, A.~Ezzir, R.~Cherkaoui, C.~Chaneac, M.~Nogues, H.~Kachkachi,
  D.~Fiorani, A.~M. Testa, J.~M. Greneche, and J.~P. Jolivet.
\newblock Surface-related properties of $\gamma$-fe$_2$o$_3$ nanoparticles.
\newblock {\em J. Magn. Magn. Mat.}, 221(1-2):63--79, 2000.

\bibitem{Fiorani:1998kx}
D.~Fiorani, J.~L. Dormann, F.~Lucari, F.~D'Orazio, E.~Tronc, and J.~P. Jolivet.
\newblock Dynamical magnetic behavior of interacting $\gamma$-fe$_2$o$_3$
  particles.
\newblock {\em Appl. Organomet. Chem.}, 12(5):381--386, 1998.

\bibitem{DormannReview}
J.L. Dormann, D~Fiorani, and E.~Tronc.
\newblock Magnetic relaxation in fine-particle systems.
\newblock {\em Advances in Chemical Physics}, XCVIII:283, 1997.

\bibitem{PhysRev.130.1677}
William~Fuller Brown.
\newblock Thermal fluctuations of a single-domain particle.
\newblock {\em Phys. Rev.}, 130(5):1677--1686, Jun 1963.

\end{thebibliography}
\end{document}